# Superconductivity in the Niobium-rich compound $Nb_5Se_4$

T. Klimczuk[1], K. Baroudi[2], J.W. Krizan[2], A.L. Kozub[1], and R.J. Cava[2]

[1] *Faculty of Applied Physics and Mathematics, Gdansk University of Technology, Narutowicza 11/12, 80-233 Gdansk, Poland*
[2] *Department of Chemistry, Princeton University, Princeton NJ 08544, USA*

**Abstract**

The niobium rich selenide compound $Nb_5Se_4$ was synthesized at ambient pressure by high-temperature solid-state reaction in a sealed Ta tube. Resistivity and heat capacity measurements reveal that this compound is superconducting, with a $T_c$ = 1.85K. The electronic contribution to the specific heat γ and the Debye temperature are found to be 18.1 mJmol$^{-1}$K$^{-2}$ and 298 K respectively. The calculated electron-phonon coupling constant $\lambda_{ep}$ = 0.5 and the $\Delta C_p/\gamma T_c$ = 1.42 ratio imply that $Nb_5Se_4$ is a weak coupling BCS superconductor. The upper critical field and coherence length are found to be 1.44 T and 15.1 nm, respectively.

Corresponding author: Tomasz Klimczuk (e-mail: tomasz.klimczuk@pg.gda.pl)

1. **Introduction**

Although it is a poor metal, niobium holds the record for the highest superconducting critical temperature ($T_c$) among elements under ambient pressure. Its remarkably high $T_c$, 9.2 K, influenced physicists to search for superconducting alloys in the 1960s and resulted in finding the now well-known NbTi and $Nb_3Sn$ superconductors. From an empirical perspective, one can therefore propose that metallic compounds containing Nb in significant proportion are good candidates for superconductivity. Of particular interest due to the fact that they can display electronic ground states that compete with superconductivity are compounds of Nb with the chalcogenide element Se. There are nine binary Nb-Se compounds reported in the literature, displaying a variety of physical properties: charge-density-wave (CDW) formation in $NbSe_2$ [1, 2] and $NbSe_3$ [3, 4], semiconducting behavior in $Nb_2Se_9$ [5], and superconductivity in $Nb_3Se_4$ [6] and $NbSe_2$ [7, 8]. The coexistence of multigap superconductivity, $T_c \sim 7K$, with CDW formation, $T_{CDW} = 35$ K [e.g. refs. 1, 2], makes $NbSe_2$, for instance, a particularly interesting material.

In contrast to the case for $NbSe_2$, there are not many reports focused on the physical properties of metal rich (i.e. a compound with more metal atoms than nonmetal atoms) niobium selenides. This is likely caused by the high temperature required for the synthesis of these compounds and the volatility of elemental Se. K. Tsukuma et. al [9] studied the $Nb_5Se_{4-x}S_x$ system synthesized at high pressures. The properties of the binary compound $Nb_5Se_4$ were not reported, but resistivity measurements on a high-pressure-synthesized sample with composition $Nb_5Se_2S_2$ showed a superconducting transition at 3.4 K. Here we report the high temperature synthesis, at ambient pressure, of the metal rich niobium selenide $Nb_5Se_4$. We find that this compound is superconducting at 1.85 K. Characterization of the superconductivity of $Nb_5Se_4$ indicates that it is a weak coupling BCS superconductor. We compare its superconducting properties to those of elemental $Nb_3Se_4$, $NbSe_2$ and $Nb_{5-\delta}Te_4$.

## 2. Experimental

A polycrystalline sample of $Nb_5Se_4$ was synthesized by high-temperature solid-state reaction from $NbSe_2$ and elemental Nb. The $NbSe_2$ precursor was prepared by heating Nb powder (Alfa Aesar 99.9%) and Se (Alfa Aesar 99.999%) in an evacuated quartz tube at 870 K for 5 hours followed by heating at 1020 K for 36 hours. The powder of the precursor was determined to be pure by the x-ray diffraction technique. $NbSe_2$ was then mixed with Nb powder in the appropriate stoichiometric ratio, thoroughly ground, pelletized under a pressure of 300 MPa, and sealed in a Ta ampoule in a high purity Ar atmosphere. The ampoule was subsequently heated at 1620 for 5 hours in a vacuum furnace (Materials Research Furnaces, Inc.). This formed the $Nb_5Se_4$ compound. To anneal out possible defects, the pellet was then heated in an evacuated sealed quartz glass tube at 1170 K for 12 hours. Powder X-ray diffraction (PXRD, Bruker D8 Focus, Cu $K_\alpha$ radiation, graphite diffracted beam monochromator) was used to structurally characterize the sample by the Rietveld method [10] through use of the FullProf 5.30 program [11]. Measurements of the temperature dependence of the electrical resistivity and heat capacity were performed in a Quantum Design Physical Property Measurement System (PPMS) equipped with a $^3$He cryostat.

## 3. Results

The room temperature powder X-ray diffraction (PXRD) pattern for the $Nb_5Se_4$ sample, with a successful structural fit of the data to the $Ti_5Te_4$ structure type, reported previously for $Nb_5Se_4$ [12] is shown in Fig. 1. The excellent quality of the refinement and the absence of additional reflections in the PXRD pattern confirm the high purity of the $Nb_5Se_4$ sample. The estimated lattice parameters and the 8$h$ position coordinates obtained in the refinement are close to those reported by Selte and Kjekshus [12]. The refined structural parameters are summarized in Table I.

A structural model with niobium vacancies present was also tested, in analogy to what is found for the nonstoichiometric compound $Nb_{5-\delta}Te_4$ (ref. 13). This model did not result in a significant change to the already excellent fit. This is commensurate with the difference in synthetic procedures for these two materials. Synthesizing $Nb_5Se_4$ as a pellet in a sealed tantalum tube does not offer a mechanism for the bulk sample to be niobium poor. The tantalum tube was clean and did not show any sign of a chemical attack, and thus the presence of an elemental niobium metal impurity, as would be required for a matter-conserving synthesis if the $Nb_5Se_4$-type phase is Nb deficient, would be readily detectable in the powder diffraction pattern of the bulk material and through the presence of a superconducting transition at the $T_c$ of Nb, neither of which is observed. Given the excellent fit of the diffraction data to the ideal structure model, and the lack of any other evidence for nonstoichiometry, the composition of the phase is well established.

$Nb_5Se_4$ has a tetragonal, body-centered crystal structure, based on square prisms of niobium atoms, with short edges of 3.14 Å in the plane and long edges of 3.45 Å along the $c$ axis. The prisms share square faces to form infinite chains along $c$ (inset of Figure 1). Each prism has another Nb atom in its center, such that the structure can be considered as consisting of chains of body centered, stretched Nb cubes sharing faces along $c$, separated by Se atoms. Alternatively, the crystal structure of $Nb_5Se_4$ can be seen as infinite chain of compressed $Nb_6Se_8$ clusters [13].

The temperature dependence of the electrical resistivity $\rho(T)$ of $Nb_5Se_4$ from 0.5K to 300 K, measured in zero applied magnetic field, is presented in the main panel of Figure 2. The $\rho(T)$ data demonstrates a positive derivative ($d\rho/dT > 0$) and a residual resistivity ratio RRR slightly less than 2, which is typical for a metal in a polycrystalline form where grain boundaries conduct poorly [14]. The red line through the experimental data represents a fit that combines a Bloch-Grüneisen resistivity $\rho_{BG}$ together with a parallel resistor $\rho_p$:

$$\rho(T)^{-1} = \rho_P^{-1} + (\rho_0 + \rho_{BG})^{-1} \text{, where}$$

$$\rho_{BG} = 4R\Theta_R \left(\frac{T}{\Theta_R}\right)^5 \int \frac{x^5}{(\exp(x)-1)(1-\exp(-x))} dx.$$

The Bloch-Grüneisen fit to $\rho(T)$ gives: $\rho_0$ = 2.8 mΩ cm, $\rho_P$ = 3.9 mΩ cm, and $\Theta_R$ = 182 K. Low temperature resistivity data measured under magnetic field from 0 T to 1 T is shown in the inset of Figure 2. In the zero field measurement, the onset of the superconducting transition is $T_{c\ onset}$ ~ 2 K and a sharp superconducting transition width of about 70 mK (using the 90% - 10% criterion) is observed. The superconducting temperature ($T_c$), defined as the temperature at which the resistivity dropped by 50% from the normal state, was estimated for each applied magnetic field. These data are shown in Figure 4 and will be discussed in the following.

The characterization of the superconducting properties of Nb$_5$Se$_4$ by specific heat is shown in Figure 3. Panel (a) shows the temperature dependence of C$_p$/T at magnetic fields from μ$_0$H = 0 to 0.8 Tesla (T). A sharp λ - type anomaly confirms the presence of bulk superconductivity in Nb$_5$Se$_4$. The superconducting transition is suppressed to ~0.7 K with the application of a field of μ$_0$H = 0.8 T. The superconducting transition under zero magnetic field is presented in panel (b). The entropy conserving construction is shown by solid lines. Based on this construction, the superconducting critical temperature (T$_c$ = 1.85 K) and the superconductivity jump (ΔC/T$_c$ = 25.8 mJ mol$^{-1}$ K$^{-2}$) were estimated. The T$_c$ obtained from the heat capacity measurements is very close to the value obtained by the resistivity method (T$_c$ = 1.79 K). The heat capacity, (C$_p$/T) versus T$^2$, measured under a magnetic field μ$_0$H = 2 T, which exceeds the upper critical field for Nb$_5$Se$_4$, is presented in Figure 3(c). The data are fitted by the formula C$_p$/T = γ + βT$^2$, with the fit result represented by a red solid line. The Sommerfeld parameter (electronic specific heat coefficient) γ = 18.1 mJ mol$^{-1}$ K$^{-2}$, and phonon specific heat coefficient β = 0.658 mJ mol$^{-1}$ K$^{-4}$ are extracted from these data. The Debye temperature estimated from β by using Θ$_D$=(12π$^4$nR/5β)$^{1/3}$ is Θ$_D$ = 298(1) K.

Knowing the Sommerfeld parameter and the Debye temperature allows for the estimation of several superconducting parameters. The normalized specific heat jump value ΔC/γT$_c$ was found to be 1.42, which is almost exactly that expected for the Bardeen-Cooper-Schrieffer (BCS) weak-coupling value (1.43). Weak coupling superconductivity for Nb$_5$Se$_4$ is also suggested from the electron-phonon constant value λ$_{ep}$ = 0.5, obtained from the inverted McMillan equation:

$$\lambda_{ep} = \frac{1.04 + \mu^* \ln\left(\frac{\theta_D}{1.45 T_C}\right)}{(1 - 0.62\mu^*)\ln\left(\frac{\theta_D}{1.45 T_C}\right) - 1.04}.$$

For this calculation we used the Coulomb repulsion constant μ$^*$ = 0.13, which falls in the range of 0.1–0.15 used in the literature. The same value was also used for example by Karki, et al. (Ref. 15). Having estimated λ$_{ep}$, the density of states at the Fermi energy (for both spin directions) was calculated by using the relation: $DOS(E_F) = \frac{6\gamma}{\pi^2 k_B^2 (1 + \lambda_{ep})}$. It was found that DOS(E$_F$)

= 5.1 states / eV f.u., very close to the reported DOS($E_F$) = 4.9 states / eV f.u. for $Nb_{5-\delta}Te_4$ (ref. 13).

Figure 4 presents the temperature dependence of the upper critical field for $Nb_5Se_4$, with the data points taken from the heat capacity (open squares) and resistivity (closed circles) measurements. The heat capacity data can be reasonably fitted by the theoretical expectations of the single band Werthamer-Helfand-Hohenberg (WHH) model in the dirty limit [16]. In this model $H_{c2}(0)$ is calculated from the formula given in terms of digamma functions $\ln\left(\frac{1}{t}\right) = \psi\left(\frac{1}{2} + \frac{\bar{h}}{2t}\right) - \psi\left(\frac{1}{2}\right)$, where $t = \frac{T}{T_c}$ is the reduced temperature and $\bar{h} = \frac{4H_{c2}}{\pi^2\left(-dH_{c2}/dt\right)_{t=1}}$. From the fit we estimate $\mu_0 H_{c2}(0) = 0.98$ T. This is not consistent with the resistivity measurement that show a $T_c = 0.55$K under applied magnetic field of 1T; thus a different analysis of this data is required.

In contrast, the temperature dependence of $H_{c2}(T)$ obtained from the resistivity measurement is almost perfectly linear ($R^2=0.9998$), with the negative slope $d(\mu_0 H_{c2})/dT = -0.812(5)$ T/K and the intercept value $\mu_0 H_{c2}(0) = 1.444(6)$ T. Such a linear behavior for $H_{c2}(T)$ is predicted for a spheroidal Fermi surface [17]. Considering both the heat capacity and resistivity measurements, it is more consistent with the present results than the WHH model. Taking $\mu_0 H_{c2}(0) = 1.444$ T, the superconducting coherence length for $Nb_5Se_4$ can be calculated by using the Ginzburg-Landau formula $\xi_{GL}(0) = \{\phi_0/[2\pi H_{c2}(0)]\}^{1/2} = 15.1$ nm. The superconducting properties of $Nb_5Se_4$ are compared to those of other niobium selenides in Table II.

4. **Conclusions**

In summary, we have successfully synthesized the niobium rich $Nb_5Se_4$ compound in high purity. Its crystal structure contains infinite chains of elongated body centered Nb cubes. In the normal state $Nb_5Se_4$ is a poor metal with RRR ~ 2 and $\rho_0 = 1.6$ mΩcm. The superconducting critical temperature, $T_c = 1.85$K, was determined by electrical resistivity and heat capacity measurements. The Sommerfeld parameter $\gamma = 18.1$ mJ mol$^{-1}$ K$^{-2}$, and the Debye temperature $\Theta_D = 298$ K for $Nb_5Se_4$ were estimated from the fit to the low temperature heat capacity data. The Debye temperature is higher than reported for $Nb_{5-\delta}Te_4$ (ref. 13) and reflects lower atomic mass of the Se atom. The normalized specific heat jump value $\Delta C/\gamma T_c = 1.42$ and the electron-phonon constant value $\lambda_{ep} = 0.5$, suggest that $Nb_5Se_4$ is a BCS-type weak coupling superconductor.

The superconducting critical temperature, $T_c = 1.85$ K, observed for $Nb_5Se_4$ is the lowest superconducting temperature among Nb-Se binary compounds. The almost doubled $T_c = 3.4$ K observed for $Nb_5Se_2S_2$ [9], which has the same crystal structure as $Nb_5Se_4$, may be partially explained by replacement of half of the selenium by much lighter sulfur atoms in $Nb_5Se_{4-x}S_x$. Similarly, the much lower $T_c$ observed for $Nb_{5-\delta}Te_4$ ($T_C = 0.6 - 0.9$ K, ref. 13) may be partially explained by the presence of the much heavier Te atom when compared to Se. Indeed, among three parameters that influence $T_c$ only the Debye temperature differs, whereas estimated $\lambda_{ep}$ and $DOS(E_F)$ for $Nb_5Se_4$ and $Nb_{5-\delta}Te_4$ are almost the same. The characterization of related materials with the same structure type would be of significant interest.

**Acknowledgments**

The materials synthesis and structural characterization performed at Princeton was supported by the US Department of Energy, grant DOE FG02-98ER45706. The research performed at the Gdansk University of Technology was financially supported by the National Science Centre (Poland) grant (DEC-2012/07/E/ST3/00584).

Tables:

TABLE I

Refined structural parameters for $Nb_5Se_4$ at 298 K. Space group I 4/m (s.g. # 87), $a = 0.985247(23)$ nm, $c = 0.345217(8)$ nm. Figures of merit: goodness of fit $\chi^2 = 1.84$, weighted profile residual $R_{wp} = 18.9$ %, profile residua; $R_p = 17.1$ %.

| $Nb_5Se_4$ | | | |
|---|---|---|---|
| Atom | Wyckoff position | x | Y |
| Nb1 | 2a (0, 0, 0) | 0 | 0 |
| Nb2 | 8h (x, y, 0) | 0.30437(16) | 0.37312(17) |
| Se | 8h (x, y, 0) | 0.05591(19) | 0.28292(19) |

TABLE II. Superconducting parameters of $Nb_5Se_4$

| Parameter | Unit | $Nb_5Se_4$ [a] | $Nb_3Se_4$ [b] | $NbSe_2$ [c] | $Nb_{5-\delta}Te_4$ [d] |
|---|---|---|---|---|---|
| $T_c$ | K | 1.85 | 2.31 | 7.09 | 0.6-0.9 |
| $\Theta_D$ | K | 298 | 220 | 236 | 259 |
| $\gamma$ | mJ mol$^{-1}$ K$^{-2}$ | 18.1 | 21.4 | 5.8 | 16.5 |
| $\Delta C/\gamma T_c$ | | 1.42 | 0.66 | 1.97 | ~1.4 |
| $\lambda_{ep}$ | | 0.5 | 0.51 | 0.79 | 0.44 |
| $DOS(E_F)$ | states / eV f.u | 5.1 | 3.0 | 1.4 | 4.9 |

a) This work;

b) ref. 18;

c) ref. 19, $\lambda_{ep}$ calculated assuming $\mu^* = 0.13$;

d) ref. 13

**Figures**

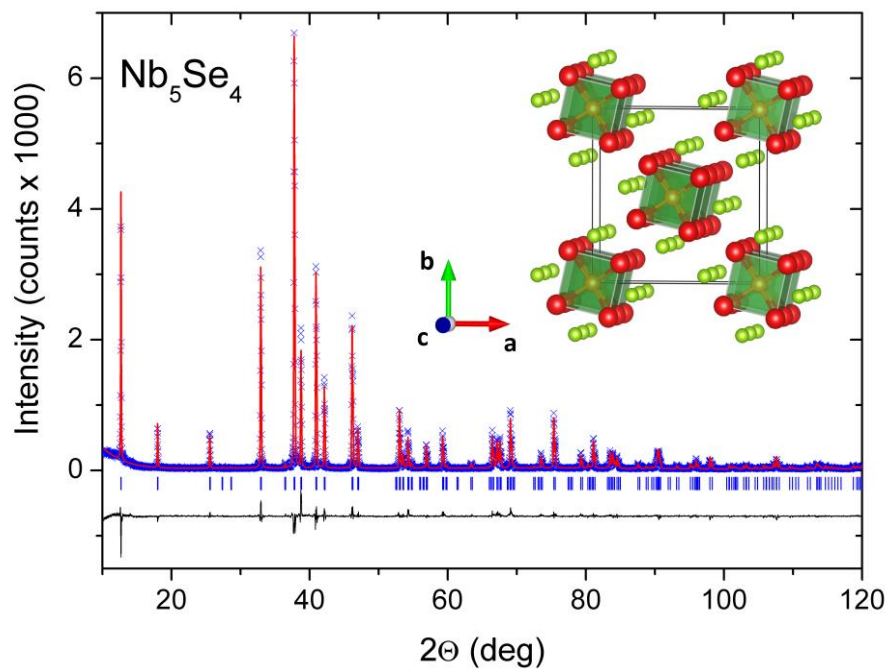

**Figure 1**

Rietveld refinement of the room temperature powder X-ray diffraction data for $Nb_5Se_4$. (Cu Kα radiation) Observed data and calculated intensity are represented by the crosses and the solid red line, respectively. The difference is shown in the lower part by a solid black line. The blue vertical ticks correspond to the Bragg peaks for $Nb_5Se_4$: space group I 4/m (s.g. # 87), $a$ = 0.985247(23) nm, $c$ = 0.345217(8) nm. The inset shows the crystal structure of $Nb_5Se_4$: the Nb ions are orange (2a site) and red (8h site), Se ions are green.

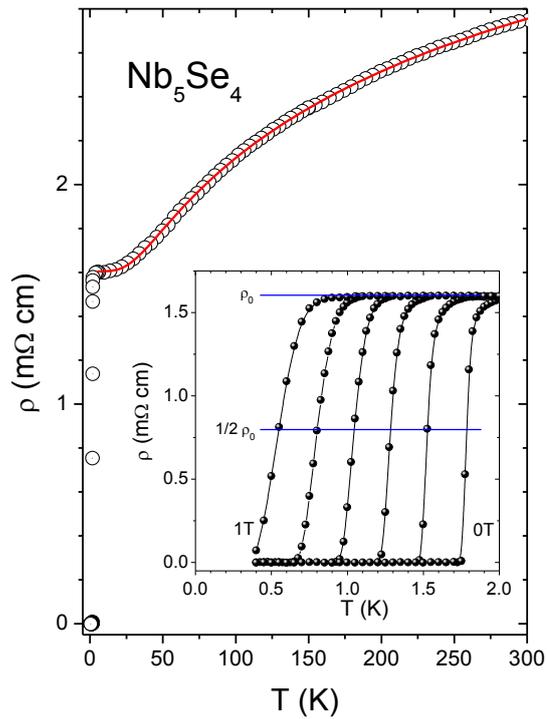

**Figure 2**

Characterization of the Nb$_5$Se$_4$ superconductor through resistivity measurements. Main panel: Electrical resistivity versus temperature for Nb$_5$Se$_4$ with applied magnetic field $\mu_0H = 0$ T. The solid line is a fit that combines Bloch Grüneisen resistivity $\rho_{BG}$ together with a parallel resistor $\rho_p$. Inset: expanded plot of $\rho(T)$ showing the superconducting transition for different values of the applied magnetic field from $\mu_0H = 0$ to 1 T.

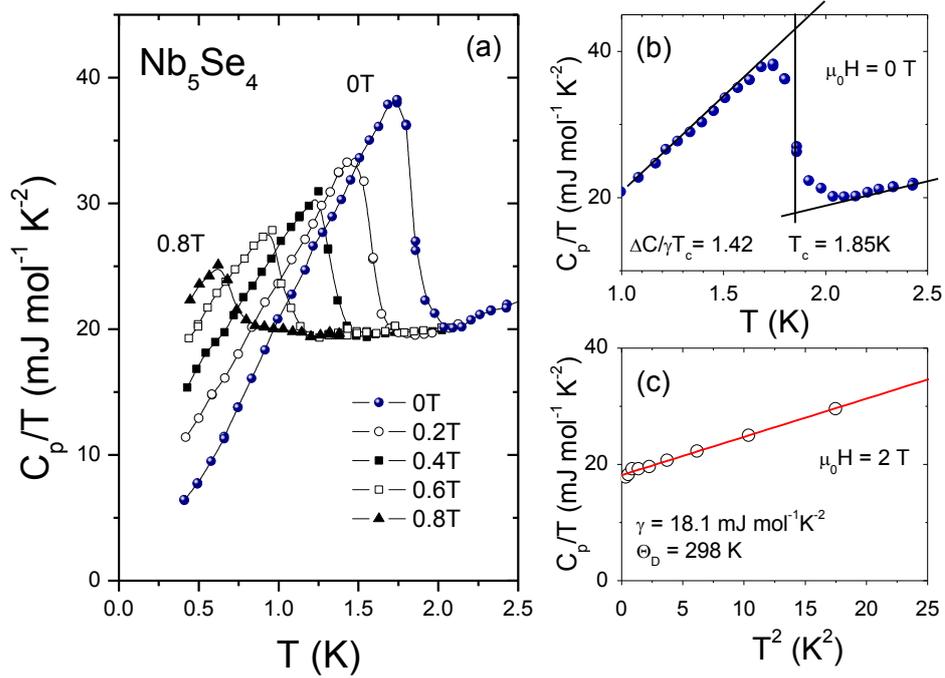

**Figure 3**

Characterization of the Nb$_5$Se$_4$ superconductor through specific heat measurements. (a) Heat capacity plotted as C$_p$/T versus T between 0.4K and 2.5K measured with various applied magnetic fields. The solid line in the inset (b) is the entropy conserving construction to estimate the heat capacity jump ΔC/T$_c$ as well as T$_c$. (c) C$_p$/T versus T$^2$ measured with an applied field of μ$_0$H = 2 T. The solid red line is fit by expression C$_p$/T = γ + βT$^2$.

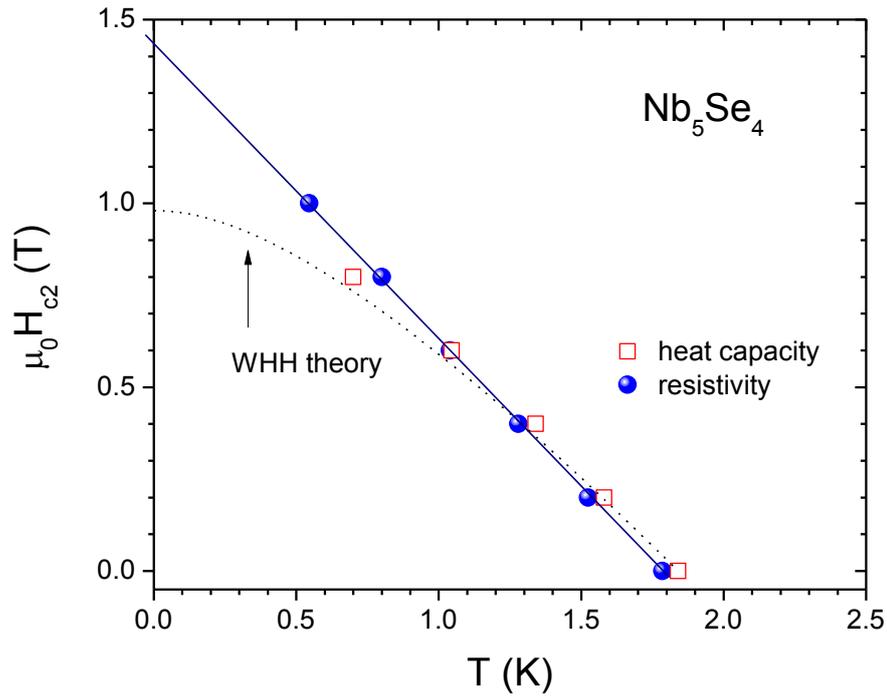

Figure 4.

The temperature dependence of the upper critical field for $Nb_5Se_4$. Open squares and closed circles are data taken from the heat capacity and resistivity measurement, respectively. The solid line shows the linear fit, whereas the dashed line is the fitting curve determined by the WHH formula.